\documentclass[10pt,conference]{IEEEtran} 

\usepackage{url}
\usepackage{mathpazo} 
\usepackage{graphicx}

\title{Enhancing Software Development with Context-Aware Conversational Agents: A User Study on Developer Interactions with Chatbots}

\author{
\IEEEauthorblockN{Glaucia Melo\IEEEauthorrefmark{1}, Paulo Alencar\IEEEauthorrefmark{2}, Donald Cowan\IEEEauthorrefmark{2}}
\IEEEauthorblockA{\IEEEauthorrefmark{1}Department of Computer Science, Toronto Metropolitan University, Toronto, Canada \\
glaucia@torontomu.ca}
\IEEEauthorblockA{\IEEEauthorrefmark{2}David R. Cheriton School of Computer Science, University of Waterloo, Waterloo, Canada \\
\{palencar, dcowan\}@uwaterloo.ca}
}

\begin{document}
\maketitle

\begin{abstract}

Software development is a demanding process that requires constant multitasking, adherence to evolving procedures, and continuous learning. With the increasing complexity of modern technology, particularly the rise of Generative AI, supporting developers in their tasks has become crucial. Although tools like OpenAI’s ChatGPT-4, Microsoft Copilot, and Claude have been introduced to assist software developers, little is known about the features developers desire in chatbots that support their work. Our study explores what features developers want in text-based conversational agents (CAs), or chatbots, that support developers' work. We conducted a study with 29 developers using a prototype chatbot, revealing significant interest in task automation and version control support. Our findings underscore the importance of deep contextual understanding and personalized assistance in these tools, particularly the need for chatbots to adapt to novice and experienced developers. Key insights from the study include the critical role of context discovery and historical interaction analysis in chatbot design, offering new perspectives on how CAs can be integrated into development workflows. This research contributes to designing context-aware chatbots to enhance productivity and developer satisfaction, address the gap in understanding developer preferences for chatbot features, and set the stage for future exploration of human-AI collaboration in software engineering.
\end{abstract}


\section{Introduction}

Developing software is a challenging task. For example, developers typically work on multiple projects, comply with company processes, demonstrate technical knowledge and soft skills, attend meetings, and train newcomers. In this very dynamic and rich environment \cite{bradley2018context, MEYER2017, Storey2022}, proposing solutions to support software development has been a topic of increasing interest both in research and industry, as developers often search and require more knowledge than they have at their immediate disposal \cite{6671289}. Moreover, as companies become more technology-centric \cite{tech_company,vial2021understanding}, facilitating the job of software developers is becoming a significant issue.

Various efforts have been explored to support software developers, from adopting new processes and tools to adhering to domain-specific practices. While developers have historically relied on online content like StackOverflow to support their work \cite{melo2019retrieving}, more specialized approaches have been proposed to assist developers in diverse ways, including artifact retrieval \cite{sawadsky2011fishtail}, software maintenance \cite{972666}, and task automation \cite{lin2020msabot}. One major challenge developers face is keeping track of task details and managing frequent context changes \cite{MEYER2017}. Solutions for these tasks can enhance productivity by reducing reliance on memory and extensive documentation. While automation of these tasks has been pursued, it introduces new challenges \cite{bradley2018context}. For instance, identifying what needs to be executed in a given process context is complex, and developers must remember specific scripts to automate their tasks. 

Considering these automation challenges, text-based contextual conversational agents (CAs) or chatbots offer an alternative approach to assist developers. Chatbots use natural language, eliminating the need for developers to learn complex command executions, and can be integrated directly into integrated development environments (IDEs) and existing tools while preserving conversation histories and allowing for high customization according to users’ needs. Chatbots have shown promise in supporting software developers in various scenarios, with case studies illustrating their use across different contexts \cite{Okanovi2020, Abdellatif2022, bradley2018context}. More recently, large language model (LLM)-powered solutions such as ChatGPT have become valuable resources for software development, providing code suggestions, debugging assistance, and quick access to general programming knowledge \cite{terzi2024using, Chouchen24}. These advancements have demonstrated potential for enhancing productivity, particularly with repetitive tasks or general coding queries.

However, a critical limitation of these tools is that they are not tailored specifically for the software development process and are not built on a foundation of developers’ specific requirements and preferences. Software development is characterized by rapid changes, evolving technologies, and complex contextual requirements that present significant challenges for generic LLMs \cite{Raiaan2024, ratner-etal-2023-parallel}. These tools still struggle with accessing real-time information, integrating into specialized development environments, and understanding the deeply contextual, project-specific nuances crucial for developers. Additionally, they often lack the adaptability needed to align with individual workflows, coding styles, or the intricate dependencies within a project’s codebase, highlighting the need for more specialized support systems in this field. Research suggests that having a human-in-the-loop can further enhance the effectiveness of such tools in software engineering by bridging some of these gaps \cite{Moreb2020, melo2019retrieving, Nascimento18SWEvsML}.

Conversational agents (CAs), a subset of AI technology, have transformed human-system interactions and present unique opportunities and challenges in software development. Research shows that CAs can transition systems from graphic to conversational interfaces, potentially increasing productivity \cite{folstad2017chatbots,brandtzaeg2017people}. Chatbots can benefit developers who lack proper technical knowledge \cite{lebeuf2017software} and assist novice developers \cite{Okanovi2020}. Studies such as PerformoBot \cite{Okanovi2020} and Liu et al. \cite{Liu2020Understanding} highlight the positive impact of chatbots on user experience and developer support. There's a need to integrate chatbots with development workflows \cite{brown2020sorry,Tonder2019}. A tool that captures the current context (artifacts, team members, project information) and guides developers through the development process (or executes tasks on developers' behalf) is yet to be evaluated. While chatbots offer a wide range of capabilities for developers, as stated, we are still questioning whether they truly meet the needs and expectations of developers.

For this goal, we proposed a series of studies with software developers to investigate (1) the extent to which developers are willing to use chatbots for task execution, (2) extract requirements from developers for such tools, and (3) support the design of conversational agents (CAs) for software development assistance. This investigation aims to provide insights into developers' interest in contextual support through CAs. Additionally, this research informs the design of context-aware conversational tools for software developers and supports the customization of systems to developers' preferences. Studying the preferences of developers using chatbots remains crucial, even with the advent of LLMs. While LLMs offer advanced capabilities, developers' needs, workflows, and expectations are diverse and continually evolving. Understanding these preferences allows for developing more targeted, efficient tools that complement LLMs, ensuring that chatbot systems align with real-world software development tasks, streamlining developer productivity, and reducing friction in the software development process. Moreover, insights into user preferences can guide the enhancement of LLMs to provide more relevant, context-aware support for software developers.

We conducted this exploratory study to understand software developers' preferences and requirements when supported by a text-based conversational agent (CA). Our goal was to investigate developer-chatbot interactions and gather requirements to inform the development of such tools. Instead of evaluating specific tools or their features, we focused on capturing interaction aspects and developers' preferences using a text-based chatbot. The investigation was carried out in two parts. First, we used a Wizard-of-Oz methodology with five participants in a classroom environment. We collected user study data and participant perceptions to refine the study design. In the second part, reported in this paper, we incorporated feedback from the first part and designed a new study. This time, we used a mixed-methods approach and interviewed 29 software developers using a chatbot prototype explicitly created to support this research methodology. Data was gathered through interviews, questionnaires, and interactions between developers and the chatbot. Finally, we analyzed the outcomes using both quantitative and qualitative methods.

By interacting with the prototype, developers experienced using a real chatbot tool and reported their experiences through questionnaires and interviews. Our specific goals with this study are to:

\begin{itemize}
\item Collect and reflect on software developers' perceptions and experiences with chatbots supporting their work during software development.
\item Report the perceived opportunities and challenges described by developers in using chatbots.
\item Gather requirements for the development of CAs in software engineering.
\item Identify potential areas for improvement of chatbots as tools to connect software developers to technical or non-technical support they need in their work.
\item Provide a qualitative study design addressing chatbots to support developers in their work.
\item Present a discussion of the study results, highlighting methods to improve the user experience of developers during software development.
\end{itemize}

This study does not aim to evaluate the prototype directly. Instead, it leverages the prototype to elicit requirements and gather participants’ perspectives on using chatbots as a development tool. By focusing on developers' experiences—rather than industry-driven agendas or executive viewpoints—the study offers valuable insights into how developers interact with chatbots when addressing development-related queries. Although data collection was conducted in 2021-2022, before the public release of ChatGPT and other chatbot-based LLMs, the findings remain highly relevant, guiding how context-aware chatbots can be enhanced and integrated into modern development environments.

This paper is organized as follows: the current Section depicts the motivation for this work and the gap in the literature. Section \ref{studydesign} lays out the design of the study, while Section \ref{results} presents the study results. Then, we promote a discussion of the results in Section \ref{discussion} and an evaluation of threats to the study's validity in Section \ref{threats}. Finally, conclude the paper in Section \ref{conclusion}.

\section{Study Design}  \label{studydesign}

To explore the interaction characteristics between software developers and chatbots, we developed a prototype to support our user study. We aimed to present a simple chatbot interface to developers, utilizing a predefined scenario to maintain the study's scope. Developers were instructed to interact with the chatbot prototype for as long as they had questions to ask or around 10 minutes and ask the chatbot any relevant questions. After, we conducted interviews to collect data on their experiences. We then analyzed these questions, along with responses from questionnaires and interviews. Specifically, we aimed to investigate (1) developers' willingness to use a chatbot to support their daily work, (2) the types of tasks the chatbot should support, (3) the types of questions chatbots should be able to answer, (4) unexpected questions or requests, and (5) developers' overall opinions on using a contextual chatbot. 

\begin{figure}[ht!]
    \centering
    \includegraphics[width=0.9\linewidth]{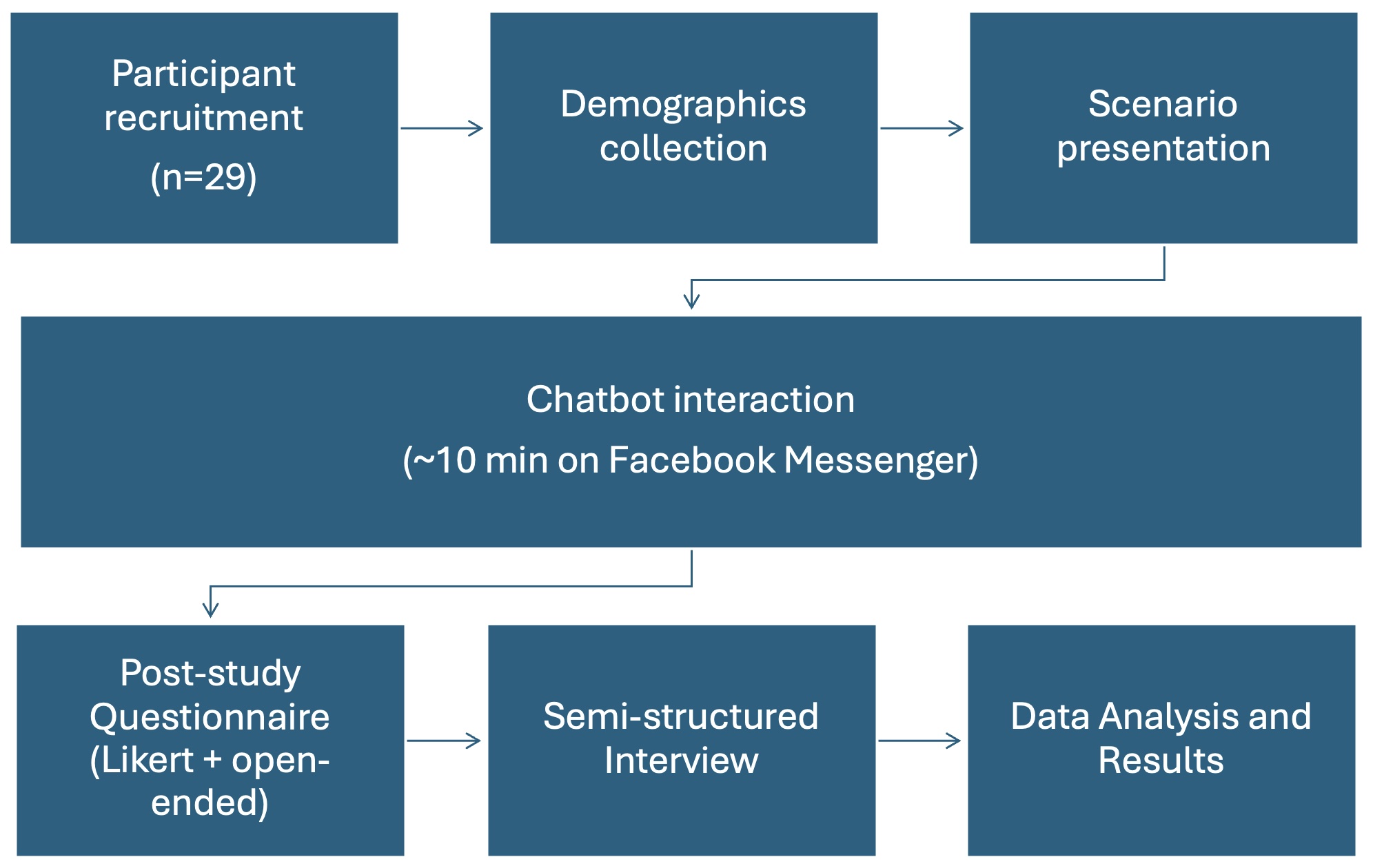}
    \caption{Software Developer Chatbot Study Design.}
    \label{fig:studydesignfig}
\end{figure}

Figure \ref{fig:studydesignfig} depicts a representation of the study design, illustrating a linear progression through seven key stages: participant recruitment, demographics collection, scenario presentation, chatbot interaction, post-study evaluation, interviews and data analysis. The central focus is the chatbot interaction phase, which occurs via Facebook Messenger for approximately 10 minutes. Following this, participants complete a questionnaire and undergo a semi-structured interview. Arrows connect each stage, showing the study's sequential nature. The diagram summarizes the research methodology, providing a visual overview of the entire study process from recruitment to final data analysis. This study methodology received ethics board clearance. 

\subsection{Prototype Design}

We developed a CA prototype to gather insights into software developers' preferences. Initially, the prototype interacted with developers through text-based chats, responding to inquiries related to software development. The interaction structure was based on a scenario presented to the participants, and the chatbot was prepared to recognize utterances related to the scenario. This scenario was validated by three experienced industry practitioners (+5 years of experience).  

\textit{\textbf{Scenario}: You arrive at your office at 8 a.m., sit on your chair and open the task manager to check what tasks are assigned. You look for the one task with the highest priority and read the text of the task. You understand what you are supposed to do for this task. You have a stand-up meeting to redefine task priority. You go to the cafeteria and grab a coffee. You’re back to your desk and open Eclipse. You sync your local code repository to get the newest code version for that specific task. You look for the artifacts you might have to edit to complete this task. You ask your colleague what he thinks about the task and the artifacts you decided to edit. You start coding the test case for that new implementation and then edit the .java class that will receive the edits. You get another coffee. You save your edits, commit your code, and create a pull request by inserting a comment in the pull request. It’s 4 pm. You update the task assigned to you to status Done and with the time you invested in solving the task. You leave the office.}

The prototype was built on the open-source Rasa chatbot platform (Rasa.com), designed for scalability and adaptability. Rasa's architecture focuses on two primary elements: NLU and dialogue management. The NLU component handles tasks like intent classification, entity extraction, and response retrieval using a trained NLU model. Dialogue management determines the following action in a conversation based on the context, guided by dialogue policies. We chose Rasa for its open-source nature, extensive documentation, and seamless integration with popular platforms like Facebook, Telegram, and Slack. This choice allowed us to deploy the chatbot on a familiar platform to software developers, ensuring a smooth user experience. Our prototype was deployed on Facebook Messenger, requiring participants to access the chatbot via Facebook Messenger. The deployment involved running the chatbot on a server and using the RASA webhook for conversation processing, with responses delivered through the Facebook app. 

\subsection{Participants}

Participants were recruited via university graduate student mailing lists, social media posts, and participant databases. Interested participants were screened based on criteria such as having at least one year of software development experience and receiving a remuneration of \$15 cad. Demographic data collected included:

\begin{enumerate}
\item Gender
\item Age
\item Highest degree or level of school completed
\item Experience with software development (in years)
\item Knowledge of software repositories (e.g., Git, Jira)
\item Previous interactions with CAs and which ones
\item Interest in CAs
\end{enumerate}

We recruited 31 participants, of whom 29 met the study criteria. Recruitment ceased once we reached saturation, meaning no new information was being added during interviews. Demographic data was collected while maintaining anonymity. Among the 29 participants, 26 had prior interactions with text-based or voice-based chatbots, and only one participant was unfamiliar with task management tools like Jira or Trello. The interview pool included 4 females and 25 males, aged 22 to 41 years. All interviews were conducted via Zoom between September 2020 and February 2021.

\subsection{Procedure Setup}

Participants joined a Zoom call and were first presented with the demographics form. After completing this form, they read the scenario describing a typical day for a software developer and were inspired by the work of Meyer et al. \cite{meyer2021todaygoodday} and validated by three industry practitioners. Participants interacted with the prototype on Facebook Messenger for about 10 minutes or as long as they had questions regarding how to support the tasks presented in the scenario. After the interaction with the chatbot, participants completed a post-study questionnaire with a Likert scale and open-ended questions:

\begin{enumerate}
\item How much did you like interacting with the chatbot?
\item How much would you be interested in using the chatbot in your company?
\item Was the chatbot helpful for your questions?
\item Did the chatbot answer what you expected?
\item Was the chatbot useful for the presented scenario?
\item How did you perceive the chatbot's response time?
\item What other steps not covered in the scenario could the chatbot be useful for?
\item What did you like about the chatbot?
\item What improvements would you suggest for the chatbot?
\item Do you think this solution adds value to software development? Why or why not?
\item Any general comments?
\end{enumerate}

Finally, we conducted semi-structured interviews asking participants to describe their experience with the chatbot. This captured opinions and insights not covered in the questionnaire. Participants turned off their cameras for the audio-recorded Zoom interviews to ensure privacy. 

Participants were unaware of the chatbot's specific capabilities to ensure they used their natural vocabulary and were not biased when asking questions. If participants asked explicitly what the bot could do or posed many unanswerable questions, the chatbot provided a list of possible actions based on the scenario. All participants were presented with the same scenario. The following section describes the study results.

\section{Study Results} \label{results}

Participants interacted with the chatbot by asking 619 questions or issuing commands, averaging 21.3 questions per participant. Each interaction lasted approximately 12 minutes, with the total interaction time amounting to 5 hours and 56 minutes. The questions were collected, and the interviews were transcribed using Whisper, an automatic speech recognition (ASR) and speech translation model\footnote{{\url{https://openai.com/blog/whisper/}}}.

\subsection{Questionnaire Data Analysis}

The post-study questionnaire provided insights into participants' experiences and perceptions of the chatbot.

\textbf{Experience interacting with the chatbot:} 
Most participants felt neutral (Likert scale = 3) about their interaction with the chatbot. Only 4 participants out of 29 reported enjoying the interaction (Likert scale = 4). Participants highlighted issues with the chatbot's ability to provide correct answers. For instance, one participant noted, \textit{"The chatbot was quick in responding but could not help me with correct answers to the questions that I had. Had it been able to answer my questions, it would have been more fun."} Another remarked, \textit{"If it can have better intelligence, I'm definitely interested in using it."} Despite these challenges, some participants appreciated its potential for automating repetitive tasks such as pushing code, committing code, creating branches, pulling/updating code, and creating pull requests.

\textbf{Willingness to use the chatbot at work:} 
Participants showed mixed interest in using the chatbot at work. Eleven participants indicated interest, while the other 18 indicated mild interest. One participant commented, \textit{"... it would be very nice to use, given a bit more access to task details."} Another highlighted its potential in training and onboarding new developers and providing quick access to task information.

\textbf{Interest in contextual information:} 
Participants recognized the importance of contextual answers when integrated with task management tools and maintaining conversation history. One participant stated, \textit{"It would also be helpful to have contextual answers, e.g., 'Start task X', and then when I'm finished I could say 'Set task completed', instead of having to remind the right task."}

\textbf{Identified design opportunities:} 
Participants expressed interest in having a chatbot that supports and leverages git functions (10 mentions), manages tasks (9 mentions), and schedules (9 mentions). They suggested the chatbot include process-related information, connect with teammates, manage tasks and schedules, and integrate with development tools like git. One participant mentioned, \textit{"I think the chatbot could be used as a developer's help tool since it clearly seemed to understand what my questions were."}

Participants noted the need to understand the chatbot's capabilities beforehand and had varied opinions on automation versus user control. Suggestions included features already available in tools like RASA, such as saving the conversation history and more complex implementations that would demand tool integration, such as managing tasks and knowing the development environment, such as for version control.

\textbf{Positive feedback:} 
Participants appreciated the chatbot's fast response time (15 mentions). They also expressed interest in features like conversation history tracking, integration with task portals, and calendar management tools. One participant remarked, \textit{"The whole idea of having an assistant to help developers manage their tasks has great potential."}

\textbf{Requirements for development of chatbots in SE:}
Participants identified several requirements for effective chatbots in software engineering, namely:
\begin{itemize}
\item Must be extremely well-designed to avoid frustration.
\item Useful for teaching beginners and integrating fragmented information/tools.
\item Adding structure to work, such as breaking down tasks.
\item Managing tasks, as navigating tools like Jira can be tedious.
\item Automating repetitive tasks for efficiency.
\item Serving as a team database/communication partner.
\item Being "smart" and capable of handling low-level tasks like managing meetings and fetching to-do lists.
\item Voice command functionality for added convenience.
\item Saving time when dealing with work-related issues outside of work hours.
\end{itemize}

One participant noted, \textit{"I think it is a great way to teach a junior/beginner developer good practices."}

\subsection{Semantic Analysis of Questions}

We performed a semantic analysis of the questions using KH Coder. The top 10 words by frequency in participants' interactions are presented in Table \ref{tab:wordfrequency2}.

\begin{table}[h]
\centering
\caption{Word Frequency in Participants' Interactions, using KH Coder.}
\label{tab:wordfrequency2}
\begin{tabular}{|l|l|l|l|l|}
\hline
\textbf{Word} & \textbf{Frequency} & & \textbf{Word} & \textbf{Frequency} \\ \hline
task & 46 & & pull (noun) & 12 \\ \hline
request & 18 & & need & 11 \\ \hline
code & 17 & & pull (verb) & 8 \\ \hline
create & 15 & & git & 7 \\ \hline
branch & 12 & & priority & 7 \\ \hline
\end{tabular}
\end{table}

Using Jaccard distance, we clustered the questions into five categories: task (24 documents), pull request/merge/branch (17 documents), code (14 documents), classes/java (14 documents), greeting (3 documents), and others (50 documents). Thirteen documents did not fit into any cluster. We consider the most frequent words to be key indicators of features that interest software developers.

\subsection{Interview Analysis}

The semi-structured interviews provided further insights into participants' interactions with the chatbot, highlighting several key themes:

\textbf{Software Development Environment Awareness:} Participants desired insights into their development environment, such as identifying artifacts to modify or colleagues' statuses and access to relevant documentation.

\textbf{Process Awareness:} Participants wanted guidance on tasks like branch naming, task status management, and accessing task-related information through the chatbot, reducing the need to navigate external portals.

\textbf{Chatbot Capabilities:} Participants emphasized the need to understand the chatbot's functionalities upfront.

\textbf{Controllability:} Participants had varying preferences for the level of automation versus control. While some favored automation, others preferred the chatbot to remain informative and guiding. Of those who expressed opinions on controllability, 9 favored automation, and 18 preferred guidance or limited automation for non-disruptive tasks like status updates.

\textbf{Chatbot Built-In Features:} Participants expressed interest in built-in features like conversation history tracking, enhancing the continuity of interactions.

\subsection{Design Opportunities - From Questionnaire}

Based on the questionnaire responses, we have identified additional feature elements for the chatbot, according to developers' responses, such as:
\begin{enumerate}
\item \textbf{Task Description and Task Portal:} Describing tasks or opening the task portal (Jira) for details, accessing external tools, integrating with tools like Jira or Trello, and providing documentation and auto-complete commands. Providing task information, changing schedules, removing tasks, and prioritizing tasks. Creating tasks and projects and providing access to related documents.
\item \textbf{Code Review and Suggestions:} Acting as a second set of eyes for code review or providing suggestions.
\item \textbf{Meeting and Appointment Scheduling: }Scheduling meetings, making appointments, and providing reminders.
\item \textbf{Developer Support:} Assisting with tasks like resolving merge conflicts and notifying about comments and events.
\item \textbf{Notifications and Aggregation:} Aggregating notifications from various sources.
\item \textbf{Git Management:} Providing information about PR reviews, comments, and status.
\item \textbf{Repository and File Navigation:} Helping locate repositories or specific files.
\item \textbf{Automating Software Development Processes:} Automating tasks like running tests and dealing with merge conflicts.
\item \textbf{Collaborative Communication and Information Sharing:} Facilitating communication between team members and providing search history.
\end{enumerate}
These responses highlight the diverse range of tasks and functionalities that participants believe the chatbot could assist with, spanning task management, code-related support, meeting scheduling, information retrieval, notifications, and collaboration.

\subsection{Design Opportunities - From Questions}

Based on the provided list of questions that participants asked the chatbot, we analyzed the most prominent themes of questions asked, obtaining further insights and design opportunities. Here are some potential categories and design opportunities for chatbots in the context of software development:

\begin{enumerate}
    \item Task Management: Participants frequently asked about their assigned tasks, priorities, deadlines, and task details. Design opportunity: A chatbot can provide users real-time information about tasks, including task names, descriptions, deadlines, priorities, and associated artifacts. It can also allow users to update task status, redefine priorities, and perform actions like marking tasks as complete or creating pull requests. An integration with project management or issue management tools is needed.
    \item Source Code and Version Control: Participants had questions about code repositories, branches, commits, pull requests, and merges. Design opportunity: A chatbot can assist users with tasks like pulling code, pushing code, creating branches, making commits, creating pull requests, merging branches, and resolving merge conflicts. It can also provide information about the status of code builds and deployments. To develop this, the implementation of integration with version control tools such as git is needed.
    \item Issue Tracking and Bug Fixes: Participants inquired the chatbot about issues, tickets, and bug fixes associated with specific tasks. Design opportunity: A chatbot can help users find related issues or tickets on platforms like GitHub or Jira. It can also provide information about the status of bug fixes, approvals from QA, and whether a build was successful.    
    \item IDE and Development Environment: Participants sought assistance with development environments, IDEs, debugging, and specific programming languages like Java. Design opportunity: A chatbot can guide IDE-related tasks, such as opening IDEs like Eclipse, helping debug code, suggesting solutions for everyday programming issues, providing syntax examples, and answering language-specific questions.
    \item Meeting and Schedule Management: Participants asked about stand-up meetings, schedules, and meeting attendees. Design opportunity: A chatbot can inform users about their schedules, upcoming meetings, attendees, agendas, and meeting locations. It can also help schedule new meetings, notify participants, and set reminders.
    \item Documentation and Artifacts: Participants requested information about documentation, requirements, and class diagrams. Design opportunity: A chatbot can assist users in finding relevant documentation, requirements, diagrams, and other artifacts associated with their tasks. It can provide links or access to these resources for easy reference.
    \item General Assistance and Miscellaneous: Participants had various general questions, such as asking for help with coding, about the weather, checking the time, requesting assistance with specific programming languages or concepts, and verifying information. Design opportunity: A chatbot can serve as a general assistant, providing help, answering questions about programming languages, explaining concepts, offering command-line examples, and performing basic tasks like checking the time.
\end{enumerate}

These categories and design opportunities can help guide the development of chatbots for software development, focusing on addressing the specific needs and workflows of developers, streamlining their tasks, and providing quick access to relevant information and resources.

\subsection{Design Opportunities - From Interviews}

Based on what participants described in the interview, the following features are desired in a chatbot:

\begin{enumerate}
    \item Participants interested in a chatbot that guides tasks and provides control:
        \begin{enumerate}
            \item Participant 2: Appreciated quick responses, prefers chatbot that guides rather than automates tasks, likes to be in control.
            \item Participant 21: Positive experience, interested in a chatbot that automates tasks and provides guidance, desires control over assigned tasks.
            \item Participant 22: Positive experience, prefers granular control over code commits, comfortable with automated commits if clear representation is provided.
        \end{enumerate}
    \item Participants struggling with chatbot usability and understanding: 
        \begin{enumerate}
            \item Participant 3: Had difficulty using the chatbot, confused about communication, prefers direct communication with task assigners.
            \item Participant 4: Found it challenging to get helpful responses, uncertain about keywords/commands, prefers chatbot that helps with specific tasks.
            \item Participant 23: Frustrated with limited capabilities and understanding of the chatbot, desires more guidance, prefers automation with safety nets.
        \end{enumerate}
    \item Participants providing general suggestions for improvement:
        \begin{enumerate}
            \item Participant 5: Recommends training the algorithm, suggests saving task IDs and conversation context, desires automation for specific tasks.
            \item Participant 19: Highlights areas for improvement, suggests handling missed tasks, understanding terminology, ontological understanding, providing task guidance, office environment integration, and task priority management.
        \end{enumerate}
    \item Participants expressing positive feedback and interest in chatbot:
        \begin{enumerate}
            \item Participant 10: Finds the chatbot useful, desires automation and voice input, and suggests integration with other systems.
            \item Participant 12: Finds the idea of a chatbot appealing and timesaving, interested in intelligence, desires guidance and automation, and asks about voice input.
            \item Participant 14: Interested in a comprehensive question-answering chatbot, prefers assistant role, desires information retrieval and code-related actions.
        \end{enumerate}
\end{enumerate}

\subsection{Demographics and Survey Correlations}

We investigated possible correlations of demographic data (age, experience) with the participants' opinions in the study. The Pearson correlation coefficient \cite{PICKARD1998811} ranges from -1 to 1, where -1 indicates a perfect negative correlation, 1 indicates a perfect positive correlation, and 0 indicates no correlation.

\textbf{Correlation Calculation:}
The correlation coefficient is a statistical measure used to quantify the strength and direction of the linear relationship between two variables. It indicates how closely the data points of these variables cluster around a linear trend. The correlation coefficient typically ranges between -1 and 1. It is important to note that the correlation coefficient measures only linear relationships. It might not capture complex relationships, outliers, or nonlinear patterns between variables. Additionally, correlation does not imply causation; a strong correlation does not necessarily mean one variable causes the other to change.

\textbf{Age x Likes the Chatbot:} To determine the correlation between age and the likeness of interacting with the chatbot, we can calculate the correlation coefficient using the study data. The correlation coefficient measures the strength and direction of the linear relationship between two variables. When comparing age and how much participants liked interacting with the chatbot, we found a correlation coefficient of -0.022. Based on this coefficient, there is a weak negative correlation (-0.022) between age and the likeness of interacting with the chatbot. One might argue that the correlation is so weak we cannot claim there is a correlation at all. 

A negative correlation means that the tendency to like the chatbot decreases slightly as age increases. However, the correlation coefficient of -0.022 indicates that this relationship is not strong. The correlation is close to zero, indicating that there is no substantial relationship between age and the likability of the chatbot. Other factors not considered in this analysis may strongly influence the likability ratings.

\textbf{Years of Experience x Likes the chatbot:} Based on the calculated correlation coefficient of 0.16, we can interpret the relationship between the years of experience and the likeness of interacting with the chatbot as follows:
\begin{itemize}
    \item The positive value of the correlation coefficient (0.16) indicates a positive relationship between the two variables. This means that, on average, as the years of experience increase, the likeness of interacting with the chatbot also tends to increase.
    \item The correlation coefficient value of 0.16 suggests a weak positive correlation. While there is a positive relationship, it is not very strong, indicating that other factors may also influence the likeness of interacting with the chatbot.
\end{itemize}

Given the weak positive correlation, we can conclude that participants with more years of experience like interacting with the chatbot slightly more than those with fewer years of experience. However, this relationship is not particularly strong, and other factors could influence the participants' preferences.

The analysis indicates a slight trend: More experienced participants tend to appreciate the chatbot more. However, the correlation is not strong enough to draw definitive conclusions. It is essential to consider other variables and conduct further research to better understand the factors contributing to the likeness of interacting with the chatbot.

\textbf{Age x Automation Preference:} Based on the provided data and the correlation analysis, the correlation coefficient between automation preference and age is approximately -0.15. This value suggests a weak negative correlation between the two variables. A negative correlation means that as age increases, there is a tendency for the preference for automation to decrease slightly. However, the correlation coefficient of -0.15 indicates that this relationship is not very strong.

It is important to note that correlation does not imply causation, and this analysis is based on a limited set of data. Therefore, it is crucial to interpret the results with caution. Other factors beyond age may also influence the preference for automation, and additional data or a more comprehensive study would be required for a more accurate analysis.

\textbf{Gender x Automation Preference:} Based on the given data, we cannot determine the correlation between automation preference and gender. Further analysis with an appropriate statistical method using a more extensive dataset would be required to assess any potential relationship between the two variables. The cross-tabulation of gender x automation preference is presented in Table \ref{tab:gender_automation_preference}. There are only 28 responses in the table because we were unable to gather the preferences of one of the participants. \textit{Yes} in the table means (prefers automation), \textit{No} means prefers guidance and \textit{Mix} means prefers a mix between automation and guidance.

\begin{table}[h]
\caption{Gender x Automation Preference.} 
\centering
\label{tab:gender_automation_preference} 
\begin{tabular}{l|l|l|}
\cline{2-3}
                          & \textbf{Male} & \textbf{Female} \\ \hline
\multicolumn{1}{|l|}{\textbf{Yes}} & 3    & 1      \\ \hline
\multicolumn{1}{|l|}{\textbf{No}}  & 5    & 2      \\ \hline
\multicolumn{1}{|l|}{\textbf{Mix}} & 14   & 3      \\ \hline
\end{tabular}
\end{table}

\textbf{Experience x Automation Preference:} When analyzing the correlation between years of experience and automation preference, results also show a preference for a mix between automation and guidance in all three experience ranges. We collected ranges of experience of $\ge$ 1 year  $\leq$ 5 years (juniors), $\ge$ 5 years $\leq$ 10 years (mid-level), and $\ge$ 10 years, seniors. 

The three categories (juniors, mid-levels and seniors) prefer a mix between automation and guidance. Of juniors, 46\% prefer a mix of automation, while 34\% prefer no automation and only guidance. Of mid-levels, 50\% prefer a mix, while 33\% prefer automation. Of seniors, 100\% of them responded they prefer a mix of full automation and guidance.

\textbf{Sentiment Analysis:} In mixed-methods research, which combines both qualitative and quantitative approaches, conducting sentiment analysis on interview results can provide valuable insights and enhance the overall depth of understanding. Sentiment analysis involves the automated process of determining the sentiment or emotional tone expressed in a piece of text, whether it's positive, negative, or neutral. By incorporating sentiment analysis, researchers can gain a deeper understanding of the emotional responses, attitudes, and perceptions of participants. This adds a layer of richness to the analysis beyond just extracting themes and patterns. Moreover, adding sentiment analysis to qualitative data analysis aids in interpretation. It helps researchers identify the tone and sentiment of participants' statements, making it easier to distinguish between strongly positive, mildly positive, neutral, mildly negative, and strongly negative sentiments. This nuanced interpretation contributes to a deeper understanding of participants' viewpoints. To analyze sentiment, we used the GPT-3.5 model. This model is not specifically dedicated to sentiment analysis like some specialized sentiment analysis models, such as the VADER (Valence Aware Dictionary and sEntiment Reasoner) model, or BERT (Bidirectional Encoder Representations from Transformers). However, it can still perform sentiment analysis by analyzing the overall tone and context of the text.

According to this model, the overall sentiment analysis is presented in Table \ref{tab:sentiment_sum}.

\begin{table}[ht!]
\caption{Sentiment Count of the User Study.}
\label{tab:sentiment_sum}
\centering
\begin{tabular}{|l|l|}
\hline
\textbf{Sentiment} & \textbf{Count} \\ \hline
Positive  & 10    \\ \hline
Negative  & 3     \\ \hline
Neutral   & 15    \\ \hline
\end{tabular}
\end{table}

\section{Discussion} \label{discussion}

We discuss some of the key implications raised by analyzing the reported results and the user study's experience.

\textbf{Topics of Interest.} Our results indicate that most of the questions asked by participants were about tasks and repository management. Of course, those were the two main topics in the scenario. However, it still shows developers are interested in receiving such support, accessing information from their tasks, or pushing code through, mostly for task management. 

\textbf{Guidance or Automation?} There are varying opinions regarding preferences between guidance and automation. This contradicts a hypothesis we had that automation would be an almost unanimous preference. Further investigations on what influences the automation preferences of users must be undertaken. 

\textbf{Need for more context.} Respondents have also indicated that adding more context to the chatbot would be desired. For example, the chatbot should know the repository address, branch, or patterns that are expected to commit messages. However, having a simple chatbot with a few expected questions and giving participants a simplwasenario were also indicated as helpful by most of the participants. Based on feedback from participants, adding context to the chatbot could immensely improve the chatbot’s capabilities, as it would be personalized. The chatbot could also incorporate the discovery of context or recommendations based on the history of interactions. With ChatGPT, many of the features that participants claimed to desire in this study have been covered. However, adding context is not one of them. Further discussion and investigations on how to add context to LLMs must be undertaken.

\textbf{Need for domain-specific knowledge.} On a similar spectrum, results also indicate that the least experienced developers account for a low score when asked if the chatbot was helpful for the questions asked. This can indicate that the chatbot has to be prepared to give more details about the process to less experienced users. Most experienced participants, who have more than five years of experience, indicated that the chatbot was more helpful.

\textbf{Chatbot acceptance.} A rather surprising result was that some participants asked questions beyond the scenario presented, which can indicate a real interest in using the chatbot in further contexts as well, not only to the ones limited to the presented scenario. Some examples include: 
\begin{itemize}
    \item More general questions such as "What's the best language to develop Machine learning models?
    \item Meeting Schedules and Agendas: Where is my next meeting? or  What is the agenda for the meeting today or When is the stand-up meeting?
    \item Code Debugging and Assistance: Can you help me in debugging code in Eclipse?
    \item Others: 
    \begin{itemize}
        \item Are we using Github or Gitlabs? or When is lunch?
        \item  Are there any updates on the Github repo while I was gone?
        \item  Can you send me the doc string for function np.argmax()?
        \item Are there any updates from the test team?
        \item Please print this document. [File attached]
\item Can you send a copy of my current edit to team X?
    \end{itemize}
\end{itemize}

\textbf{Chatbot understanding.} One negative result reported is related to how the chatbot understands the questions. Because only around 7 intentions were mapped to answers, when the developers asked questions that the chatbot did not understand but were still related to the scenario, some disappointment was reported. Although the purpose of the study was not to evaluate the tool, our prototype was very limited in that sense. Having a chatbot that could understand several other intents might have generated a more positive reaction from the developers after the chatbot interaction.

\textbf{Scenario. } Presenting participants of a mixed-methods user study with a scenario can have both advantages and disadvantages. Here's an overview of these points. We can cite as advantages:
\textbf{Contextual Understanding:} A scenario provides participants with a clear context and background for the study. This helps them understand the purpose, scope, and objectives of the research more effectively. \textbf{Engagement:} Scenarios can make the study more engaging by presenting a relatable situation. Participants can connect better with the study and its goals, which might increase their involvement and motivation. \textbf{Realism:} By presenting a scenario, you create a realistic setting that participants can relate to. This can lead to more authentic responses and insights as participants engage with the scenario as they would in a real-world context. \textbf{Consistency:} When all participants start with the same scenario, it ensures a consistent starting point for the study. This reduces variability in participants' initial understanding and sets the stage for more meaningful comparisons. \textbf{Guided Exploration:} Scenarios guide participants toward specific topics or aspects of interest. This helps ensure that all participants explore the same core concepts, making the study's findings more focused and relevant.

We can cite as disadvantages: \textbf{Bias and Stereotyping:} Scenarios could unintentionally introduce bias or stereotyping, influencing participants' perceptions or responses. Careful crafting of scenarios is needed to avoid any unintended implications. We have crafted the scenario with software developers and with the support of relevant literature \cite{meyer2021todaygoodday}. \textbf{Limited Flexibility:} Some participants might feel constrained by the scenario and may not be able to express their genuine thoughts that fall outside its scope. \textbf{Artificiality:} Depending on the complexity of the scenario or the experience of participants, they might feel that the scenario is artificial or detached from their real experiences. This could affect the authenticity of their responses.
\textbf{Misunderstanding:} There is a risk that participants might misunderstand the scenario or its intent. This can lead to participants providing irrelevant or inaccurate responses that don't align with the study's goals.
\textbf{Limited Generalization:} While scenarios provide a focused context, the findings might not be as generalizable to broader contexts, as participants are responding to a specific situation.

We believe that understanding the interaction of developers with the systems as chatbot users is key to improving developers' experience and advancing software engineering practices, providing the needed timely support for developers. Nonetheless, it seems that even having a chatbot that is limited in terms of what it can do and how it can help, participants were keen on the idea of using a chatbot. These empirical findings are aligned with the discussions presented in \cite{DagstuhlSeminarBots}. It is worth mentioning this analysis was executed in 2021 before ChatGPT was available to the public.

\section{Threats to Validity} \label{threats}

This section addresses potential threats to the validity of the user study conducted to investigate software developers' interactions with the chatbot prototype.

\subsection{Construct Validity}
\textbf{Scenario Realism:} The study scenario, depicting a typical day in the life of a software developer, may not fully represent the diversity of tasks and situations developers encounter. This could affect the realism of participants' interactions with the chatbot. To mitigate this, we have validated the scenario with three experienced software developers, and we also drew inspiration from Meyer et al. \cite{meyer2021todaygoodday}. \textbf{Question Prompting:} Participants were asked to generate questions for the chatbot. The framing of this request may have influenced the types of questions asked, potentially leading to biased or limited question sets. \textbf{Question Clarity:} The clarity of participants' questions may vary, potentially affecting the quality of interactions and the chatbot's ability to provide relevant responses.

\subsection{Internal Validity}

\textbf{Limited Interaction Time:} Participants were allotted a fixed interaction time of approximately 10 minutes with the chatbot. This constraint might not fully capture the potential benefits or limitations of the chatbot over extended periods of use. Most participants, though, indicated they had run out of questions to ask after around 10 minutes of interaction with the chatbot. Anyone who could think of more questions was encouraged to continue to ask. \textbf{Prior Chatbot Experience:} Participants had varying degrees of experience with chatbots. This prior experience might have influenced their expectations and assessments of the chatbot's performance.

\subsection{External Validity}

\textbf{Participant Pool:} The study recruited participants primarily from university graduate student mailing lists, potentially limiting the generalizability of the findings to a broader population of software developers. The study involved 29 participants, which may not fully capture the diversity of software developers in terms of experience, expertise, and preferences. \textbf{Platform Dependency:} The study deployed the chatbot on the Facebook Messenger platform. This choice of platform might not represent all the platforms and tools developers use in their daily work. However, deploying the same Rasa chatbot using other platforms, such as WhatsApp, Telegram, Slack, Twilio, Google Hangouts, Cisco Webex, and others is possible.

\subsection{Conclusion Validity}
\textbf{Questionnaire Subjectivity:} The Likert scale questions in the post-study questionnaire and open-ended responses are subject to participants' subjectivity and potential bias in their evaluations. \textbf{Interviewer Bias:} The study organizers conducted semi-structured interviews, introducing the possibility of interviewer bias in interpreting participants' responses.

\section{Conclusion and Future Work} \label{conclusion}

Our user study explored software developers' preferences when interacting with a chatbot, revealing several insights that hold substantial implications for human-machine interaction within the software development domain. Developers typically engage with myriad tools in dynamic environments, making it crucial to support them effectively to maintain work quality. Chatbots have the potential to enhance productivity, reduce training time, and make implicit preferences explicit through capture processes. We identified numerous design opportunities for chatbots in software engineering, demonstrating developers' willingness to engage with these tools and providing a foundation for future improvements in chatbot behavior and integration. Our study uncovered that most questions posed by participants centred around task and repository management, reflecting the core themes of the provided scenario. This indicates developers' keen interest in receiving support from CAs in these areas, particularly in task management. Using a scenario in our study offered benefits such as contextual understanding, engagement, and guided exploration; however, it also introduced potential biases, limited flexibility, and artificiality, necessitating careful scenario design and consideration of the proposed scenario's impact. Additional work is required to involve a larger population of software developers, mitigate potential generalization biases, and run the study now that LLMs have resolved some of the language problems mentioned as desired features by our research. Further investigation can examine the specific support that such a tool should provide to developers and the integration of a chatbot with a context model that understands the software development workflow and that developers can rely on to answer questions such as "Who in my team has solved a problem similar to ... "? Future work can also focus on considering personas in ChatGPT and investigating the experience of software developers at work using these tools, as well as the quality of the software created as developers rely on these tools to aid their work. 

\bibliographystyle{IEEEtran} 
\bibliography{main.bib}

\end{document}